\documentclass[preprint]{aastex7}
\usepackage{graphicx}
\usepackage{txfonts}
\usepackage{color}
\usepackage{comment}

\newcommand{\water}{H$_2$O}

\begin{document}

\title{Anatomy of Empirical Transit Spectra of Mars based on TGO/NOMAD}

\author[0000-0001-6727-125X]{Shohei Aoki}
\altaffiliation{These authors contributed equally to this work.}
\affiliation{Department of Complexity Science and Engineering, Graduate School of Frontier Sciences, The University of Tokyo, 5-1-5 Kashiwanoha, Kashiwa, Chiba 277-
8561, Japan}
\affiliation{Department of Geophysics, Graduate School of Science, Tohoku University, Sendai, Miyagi 980-8578, Japan}
\email{shohei.aoki@edu.k.u-tokyo.ac.jp}

\author[0000-0002-2786-0786]{Yuka Fujii}
\altaffiliation{These authors contributed equally to this work.}
\affiliation{Division of Science, National Astronomical Observatory of Japan, 2-21-1 Osawa, Mitaka, Tokyo 181-8588, Japan}
\affiliation{Graduate Institute for Advanced Studies, SOKENDAI, 2-21-1 Osawa, Mitaka, Tokyo 181-8588, Japan}
\affiliation{Department of Earth and Planetary Science, University of Tokyo, 7-3-1 Hongo, Bunkyo-ku, Tokyo 113-0033, Japan}
\email{yuka.fujii.ebihara@gmail.com}

\author[0000-0003-2064-2863]{Hideo Sagawa}
\affiliation{Faculty of Science, Kyoto Sangyo University, Kamigamo Motoyama, Kita-ku, Kyoto 603-8555, Japan}
\email{test}

\author[0000-0002-2662-5776]{Geronimo L. Villanueva}
\affiliation{NASA Goddard Space Flight Center, 8800 Greenbelt Rd., Greenbelt, 20771 MD, USA}
\email{test}

\author[0000-0003-3887-6668]{Ian Thomas}
\affiliation{Royal Belgian Institute for Space Aeronomy, 3 Avenue Circulaire, 1180 Brussels, Belgium}
\email{test}

\author{Bojan Ristic}
\affiliation{Royal Belgian Institute for Space Aeronomy, 3 Avenue Circulaire, 1180 Brussels, Belgium}
\email{test}

\author[0000-0001-7433-1839]{Frank Daerden}
\affiliation{Royal Belgian Institute for Space Aeronomy, 3 Avenue Circulaire, 1180 Brussels, Belgium}
\email{test}

\author[0000-0002-7989-4267]{Miguel Angel López-Valverde}
\affiliation{Instituto de Astrofisica de Andalucia, Glorieta de la Astronomia, 18008 Granada, Spain}
\email{test}

\author[0000-0002-8223-3566]{Manish R. Patel}
\affiliation{Department of Physical Sciences, The Open University, Milton Keynes, MK7 6AA, UK}
\email{test}

\author[0000-0002-4603-4009]{Jonathon Mason}
\affiliation{Department of Physical Sciences, The Open University, Milton Keynes, MK7 6AA, UK}
\email{test}

\author[0000-0001-7813-4605]{Yannick Willame}
\affiliation{Royal Belgian Institute for Space Aeronomy, 3 Avenue Circulaire, 1180 Brussels, Belgium}
\email{test}

\author[0000-0003-0867-8679]{Giancarlo Bellucci}
\affiliation{Istituto di Astrofisica e Planetologia Spaziali, Via del Fosso del Cavaliere 100, Roma, Italy}
\email{test}

\author[0000-0001-8940-9301]{Ann Carine Vandaele}
\affiliation{Royal Belgian Institute for Space Aeronomy, 3 Avenue Circulaire, 1180 Brussels, Belgium}
\email{test}

%\date{Received September XX, XXXX; accepted March XX, XXXX}

\begin{abstract}

Transit spectroscopy is a powerful tool for probing atmospheric structures of exoplanets. 
Accurately accounting for the effects of aerosols is key to reconstructing atmospheric properties from transit spectra, yet this remains a significant challenge. To advance this effort, it is invaluable to examine the spectral features of well-characterized planetary atmospheres. Here, we synthesize empirical transit spectra of Mars across different seasons based on data from the NOMAD's Solar Occultation channel onboard ExoMars/TGO, which operates at wavelengths of 0.2--0.65 and 2--4~\micron{}. 
In the generated empirical transit spectra, the atmosphere below 25 km is found to be largely opaque due to the presence of micron-sized dust and \water{} ice clouds, both of which substantially weaken spectral features.
The spectra exhibit CO$_2$ absorption features at 2.7--2.8~\micron{} and signatures of sub-micron-sized mesospheric \water{} ice clouds around 3.1~\micron{}, accompanied by a continuum slope. 
The amplitudes of these spectral features are found to vary with the Martian seasons, where the dust storms weaken the CO$_2$ signatures and strengthen the \water{} ice features, which serve as potential indicators of a dusty planet like Mars.
If TRAPPIST-1f possessed a Mars-like atmospheric structure, both CO$_2$ and \water{} ice features would be detectable at a noise level of 3~ppm, a level likely beyond current observational capabilities. 
Nevertheless, the 3.1~\micron{} feature produced by sub-micron-sized mesospheric \water{} ice clouds offers a novel avenue for characterizing the atmospheres of habitable-zone exoplanets.
\end{abstract}

%% Keywords should appear after the \end{abstract} command. 
%% The AAS Journals now uses Unified Astronomy Thesaurus concepts:
%% https://astrothesaurus.org
%% You will be asked to selected these concepts during the submission process
%% but this old "keyword" functionality is maintained in case authors want
%% to include these concepts in their preprints.


%% From the front matter, we move on to the body of the paper.
%% Sections are demarcated by \section and \subsection, respectively.
%% Observe the use of the LaTeX \label
%% command after the \subsection to give a symbolic KEY to the
%% subsection for cross-referencing in a \ref command.
%% You can use LaTeX's \ref and \label commands to keep track of
%% cross-references to sections, equations, tables, and figures.
%% That way, if you change the order of any elements, LaTeX will
%% automatically renumber them.
%%
%% We recommend that authors also use the natbib \citep
%% and \citet commands to identify citations.  The citations are
%% tied to the reference list via symbolic KEYs. The KEY corresponds
%% to the KEY in the \bibitem in the reference list below. 

%______________________________________________
\section{Introduction}
\label{s:intro}
Characterization of rocky planets is one of the central interests in exoplanet research. It is essential for both comparative planetology and astrobiological investigations beyond the Solar System. 
A promising method for studying their atmospheres is transit spectroscopy. 
The James Webb Space Telescope (JWST) has begun performing transit spectroscopy of exoplanets smaller than Neptune. 
It has detected several atmospheric species, including CH$_4$ and CO$_2$, in cool sub-Neptunes, and has even hinted at the presence of some unexpected species \citep{HolmbergMadhusudhan2024,Madhusudhan+2025}. 
Transit spectroscopy of Earth-sized planets around low-mass stars has also been attempted, although molecular signatures have not been clearly constrained \citep[e.g.,][]{Lim+2023,Barclay+2025}. 
In the near future, next-generation ground-based extremely large telescopes are also expected to carry out the search for atmospheric species in Earth-sized planets orbiting nearby low-mass stars. 

Nevertheless, reconstructing the properties of an exoplanetary atmosphere from its transit spectrum remains a challenge. 
While technical limitations such as low signal-to-noise ratios, limited spectral resolution, and restricted wavelength coverage pose clear obstacles, fundamental difficulties persist even under idealized, noise-free conditions. In particular, the impact of liquid/solid particles (e.g., clouds, aerosols, and dust) on transit spectra, as well as the interpretation of their spectral features, remains a challenge in exoplanetary characterization. 
The abundance and properties of the condensates, along with those of their gaseous precursors, can vary both spatially and temporally through complex interactions with atmospheric dynamics, making it difficult to model them self-consistently and to incorporate their effects into data analysis. 
One-dimensional atmospheric models that are used to simulate or interpret transit spectra often simplify these complexities. 
A substantial number of previous works that provided modeled transit spectra of a variety of rocky planets have typically neglected these condensates or assumed ad hoc properties for them, which may not adequately capture the diversity of real atmospheres. 
More recently, transit spectra based on three-dimensional general circulation models have also been modeled \citep[e.g.,][]{Kopparapu+2017,Fujii+2017,Fauchez+2019}. However, due to uncertainties associated with the coarse-graining of the cloud particle properties (i.e., size distribution and shape), subgrid-scale processes, and their complex interactions with radiation, it remains uncertain to what extent these models represent reality. 

In this context, studying the well-characterized atmospheres of planets in the Solar System provides invaluable benchmarks. 
In particular, solar occultation measurements conducted by recent planetary exploration missions offer atmospheric transmission data across different altitudes, which have been conveniently used to construct empirical transit spectra of Solar System planets. 
For example, the empirical transit spectra of Earth have been reconstructed from the solar occultation data obtained with the Atmospheric Chemistry Experiment Fourier Transform Spectrometer (ACE-FTS) onboard the SCISAT satellite \citep{SCHREIER2018,MacdonaldCowan2019}.
\citet{Doshi+2022} compared cloud-free and cloudy ACE-FTS/SCISAT data, highlighting the importance of aerosols. 
They showed that stratospheric clouds increase the transit depth by $\sim$10~ppm for an exo-Earth in the TRAPPIST-1 system, which suppresses absorption features of transit spectra. However, this effect does not significantly hinder JWST's ability to characterize the atmospheres of Earth-like exoplanets. 

The role of aerosols has also been explored using Titan as a testbed. 
\citet{Robinson2014} generated Titan's transit spectra in the 0.88--5\,\micron{} spectral range based on data from the Visual and Infrared Mapping Spectrometer onboard the Cassini spacecraft. 
Their results demonstrated that high-altitude hazes significantly limit the atmospheric depths that can be probed via transit spectroscopy, restricting observations to pressure levels between 0.1 and 10~mbar.
\citet{Changeat+2025} analyzed the same transit spectra of Titan using a retrieval framework and found that multiple aerosol components were required to achieve a successful fit.

This study focuses on Mars, which is the second most studied planet in terms of spectroscopic observations, after Earth. 
Mars has a tenuous atmosphere (6.1~mbar at the surface) with little atmospheric water vapor ($\sim $150~ppm) \citep[e.g.,][]{Smith2008}. 
A distinctive characteristic of Martian atmosphere is the presence of micron-sized airborne dust particles, which significantly influence the planet’s climate by modulating solar radiation and altering the atmospheric thermal structure. 
Similar to previous works, we aim to obtain empirical transit spectra based on the actual data from ExoMars Trace Gas Orbiter (TGO) mission.
While Martian atmospheric structure has been modeled by General Circulation Models (GCMs), 
which have been extensively validated against observations, GCMs are not necessarily perfect, particularly in their treatment of clouds and hazes. 
For example, \citet{Stcherbinine+2022} compared TGO water-ice cloud vertical profiles with simulations from the Mars Planetary Climate Model (PCM) and found that, while the model reproduces the overall altitude structure, it tends to underestimate cloud-top heights. Our observation-driven approach therefore provides a more direct, model-independent insight.
We note that the present atmospheric state of Mars is the outcome of its unique formation and evolutionary history, and thus may not necessarily apply to exoplanets in other systems.
However, it serves as a valuable example of a planet residing within the conventional habitable zone while being markedly different from an Earth-like habitable environment.
Characterizing the transit spectra of Mars would offer important insights into the observable signatures of dusty and dry worlds.

The organization of this manuscript is as follows: the method for generating empirical transit spectra and the resultant spectra are presented in Section 2.
The infrared portion of the spectra, which is most relevant to exoplanet observations, is further analyzed in Sections 3 and 4. 
In Section 3, we interpret the empirical spectra using forward modeling with an established radiative transfer code, guided by existing knowledge of the Martian atmosphere.
Section 4 presents more agnostic retrieval analyses, treating the spectra as if they were obtained from an exoplanet, to evaluate the potential for characterizing Mars-like exoplanets.
In Section 5, we further discuss the implications of the two properties of Mars transit spectra, time variability and the 3.1~\micron{} \water{} ice feature, in the context of exoplanet characterization. 
Finally, Section 6 summarizes the key findings and concludes. 

%______________________________________________
\section{Empirical transit spectra }
\label{s:data}

\subsection{Dataset and Method for Generating Empirical Transit Spectra}
We reconstruct the empirical transit spectra of Mars using data from the Nadir and Occultation for Mars Discovery (NOMAD) spectrometer onboard ExoMars/TGO, which is specifically designed for solar occultation measurements \citep{Vandaele2018}. In particular, we utilize data from the Solar Occultation (SO) \citep{Neefs+2015} and Ultraviolet and VISible spectrometer (UVIS) \citep{Patel+2017} channels of the NOMAD instrument. 
As the spacecraft orbits Mars, the Sun as seen by NOMAD moves vertically through the Martian atmosphere, allowing the instrument to measure transmitted sunlight at different tangential altitudes.
The SO channel covers wavelengths of 2--4~\micron{} with a spectral resolving power of $\mathcal{R}=17{,}000$ and a spectral resolution of $\Delta\lambda= $0.119--0.235~nm, whereas the UVIS channel spans 0.2--0.65~\micron{} with $\mathcal{R}=100$--500 and a spectral resolution  of $\Delta\lambda= $1.2--1.6~nm.

\begin{figure}
\centering
\includegraphics[width=0.8\hsize]{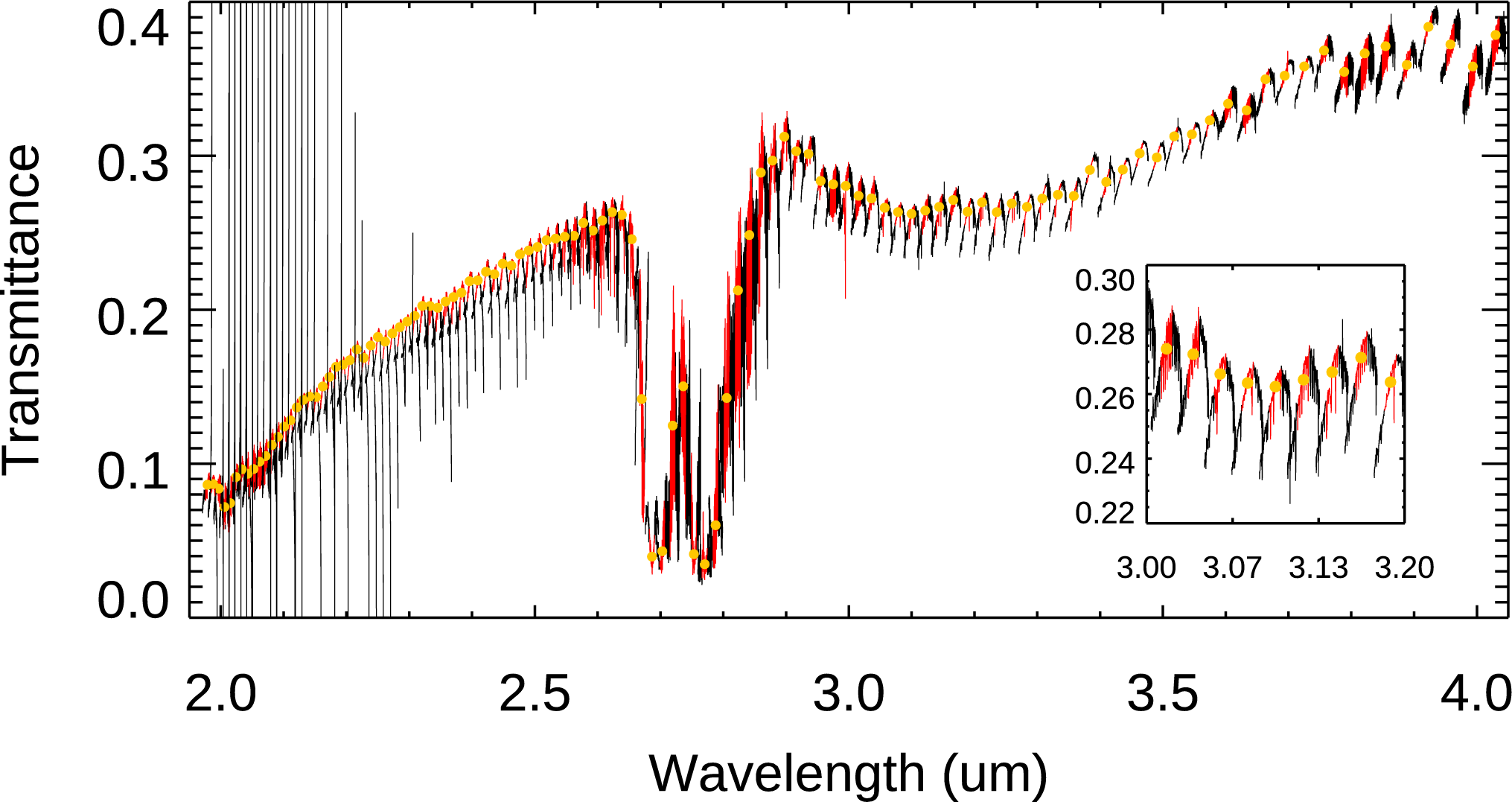}
\caption{An example of spectra obtained by NOMAD/SO in full-scan mode. The spectra were taken between 19.86 km and 22.43 km above the surface of Mars. The red curves indicate the central 120 pixels of each spectrum, which are averaged to produce empirical spectra with low spectral resolution. The averaged transmittances for each diffraction order are shown as orange dots. The inset at the lower right shows a zoomed-in view of the 3.0–3.2 \micron{} region.}
\label{Figure 1}
\end{figure}

In the SO channel, an Acousto-Optical Tunable Filter (AOTF) serves as a passband filter, selecting specific diffraction orders from the echelle grating, which provides a spectral coverage of approximately 22 cm$^{-1}$ per order. 
The AOTF allows for instantaneous switching between diffraction orders, and during standard operations, the SO channel typically records five to six orders per measurement. Occasionally, the NOMAD-SO operates in a specialized ``full-scan'' mode, in which it acquires data across 111 diffraction orders from 2 to 4 \micron{}. In this study, we analyze 29 full-scan observations obtained between 2018 and 2024, covering various locations and seasons on Mars. 
These observations were selected based on the criterion that full spectra covering 2 to 4 \micron{} were obtained at intervals of at least every 20 km within the 0-100 km altitude range. 
Figure \ref{Figure 1} shows an example of spectra obtained in the full-scan mode of NOMAD/SO. 
Each diffraction order appears as a region between two adjacent spiky features. 
These spiky features are likely due to optical modulations occurring during occultation and are subsequently corrected by our data processing routines. 
As described above, data  corresponding to different diffraction orders were acquired at different altitudes (or, equivalently, impact parameters). 
In this particular spectrum, the data points near 2~\micron{} correspond to an altitude of 19.86 km, while those near 4~\micron{} correspond to an altitude of 22.43 km; the overall increase in transmittance toward longer wavelengths is partly attributed to this variation in altitude. At higher altitudes, the transmittance generally increases because of the reduced air-mass density and optical depth.

In contrast to the SO channel, the UVIS channel continuously covers its entire spectral range using 1024 pixels, though onboard averaging is applied over 8 pixels along the wavelength axis after March 2019. The UVIS channel typically operates alongside the SO channel, and such concurrent measurements are available for 25 out of the 29 full-scan observations. 
It is important to note that UVIS data are affected by stray light contamination; however, the correction methods developed have been shown to be highly effective at mitigating this issue for wavelengths longer than 220 nm \citep{Willame+2022, Mason+2022}. Consequently, data at shorter wavelengths are excluded from our analysis.

In this paper, the empirical transit spectra refer to either the wavelength-dependent transit depth, $\Delta_{\lambda }$, or the corresponding wavelength-dependent effective atmospheric thickness, $h_{\lambda}$. 
The transit depth ($\Delta _{\lambda }$) is simply the fractional decrease in stellar flux when a planet passes in front of its host star, often represented in units of parts-per-million (ppm). 
The effective atmospheric thickness ($h _{\lambda }$), introduced by \citet{KalteneggerTraub2009} and often used in presenting modeled transit spectra,
represents the wavelength-dependent thickness of an opaque atmosphere which would result in the same transit depth, i.e.,
\begin{equation}
\Delta_{\lambda } = \frac{(R_{\rm p} + h_{\lambda})^2}{R_{\star}^2},
\end{equation}
where $R_{\rm p}$ is the planetary radius, and $R_{\star}$ is the stellar radius. 
In reality, the atmospheric transmittance gradually changes with the radius (altitude). 
Denoting the wavelength-dependent transmittance of the optical path with the impact parameter $b$ by $T(b,\lambda)$, transit depth can also be written as
\begin{equation}
\Delta _{\lambda } = \left(\frac{R_{\rm p}}{R_{\star}}\right)^2 + \frac{2}{R_{\star}^2} \int_{R_{\rm p}}^{R_{\star}} b \left[1 - T(b,\lambda)\right]\, db. 
\end{equation}
Combining these two expressions yields
\begin{equation}
h_{\lambda} = -R_{\rm p} + \sqrt{R_{\rm p}^2 + 2 \int_{R_{\rm p}}^{R_{\star}} b \left[1 - T(b,\lambda)\right] db}.
\end{equation}
For atmospheres that are much thinner than the planetary radius, as is the case here, this can be approximated by
\begin{equation}
h_{\lambda} = \frac{1}{R_{\rm p}} \int_{R_{\rm p}}^{R_{\rm top}} b \left[1 - T(b,\lambda)\right] db, \label{eq:h_eff}
\end{equation}
where $R_{\rm top}$ is the radius at the top of the atmosphere. 
In this study, the transmittance,  $T(b,\lambda)$, is derived from the NOMAD observations, and we adopt $R_{\rm top}=100$~km. 

The transmittance spectrum as a function of impact parameter, $T(b, \lambda )$, is a key input from the NOMAD measurements. However, it is not trivial to incorporate the NOMAD data into $T(b,\lambda)$, because the actual NOMAD measurements constitute a non-systematic collection of fragmented data, each taken within a narrow wavelength range (in the case of the SO channel), at a specific altitude, and at a specific location on the planet. The raw data are also affected by a signal bias due to the AOTF, which is particularly prominent near the edges of individual diffraction orders (Figure \ref{Figure 1}). To construct representative empirical transit spectra from these measurements, the following procedures are applied:
\begin{enumerate}
\item Average the transmittance spectra over each spectral interval to minimize the bias (applicable only to SO channel data);
\item Interpolate the transmittance profiles along altitude to achieve regular vertical sampling (for SO channel data, this step is performed separately for each spectral interval);
\item Compute the $h_{\lambda}$ spectrum at each observed latitude using Equation \ref{eq:h_eff}, based on the interpolated transmittance profiles;
\item Classify the set of $h_{\lambda}$ based on Martian season and latitude;
\item Average the calculated $h_{\lambda}$ spectra over latitude for each Martian season.
\end{enumerate}

In Step 1, the bias observed in the SO channel data is corrected. 
As shown in Figure \ref{Figure 1}, the spectrum of a given diffraction order displays a curved continuum, likely due to optical modulation. 
Among the 320 data points in each order, the first and last 100 pixels exhibit significant curvature.
To minimize their impact, we use only the central 120 pixels of each spectrum (highlighted in red in Figure 1) and compute the mean transmittance. This processing results in spectra with relatively low spectral resolution ($\mathcal{R} \sim 140$) where each data point is associated with an altitude (or impact parameter). Note that gaps are present between diffraction orders in the spectral coverage, which may introduce errors when comparing with radiative transfer simulations that assume continuous spectral data.

The second step (Step 2) is to obtain the transmittance spectrum as a function of altitude. 
For a given wavelength range, transmittance values as a function of impact parameter are linearly interpolated along altitude at 1 km intervals, spanning from 0 to 100 km. 
Figures 2a, 2b present an example of the resulting NOMAD/SO transmittance as a function of altitude.
In panel 2a, the dotted curves represent spectra averaged over each diffraction order, while the solid lines indicate the interpolation. Panel 2b displays the transmittance at different altitudes and wavelengths by the color gradients. 
This highlights that the lower atmosphere appears opaque due to the presence of aerosols and molecular absorption. 
Similar to the SO channel data reduction, the UVIS data are also interpolated along the altitude. 
Figures 2c, 2d present an example of the NOMAD/UVIS spectrum.

The third step (Step 3) is to compute the effective height using Equation 1.

Following this, Step 4 involves classifying the dataset based on Martian season and latitude.
It is important to note that Mars experiences seasons, similar to Earth, considering that both planets have a similar obliquity. As illustrated in Figure 3, Solar longitude (Ls) values of 0° and 180° correspond to the northern spring and autumn equinoxes, respectively, while Ls values of 90° and 270° correspond to the northern and southern summers, respectively. During the northern summer, the distance between the Sun and Mars is relatively large, resulting in lower atmospheric temperatures. This leads to a higher abundance of \water{} ice clouds in the upper troposphere and a lower presence of dust \citep[e.g.,][]{Smith2008}. Therefore, this season is sometimes referred to as the ``non-dusty period''. In contrast, the southern summer is known as the ``dusty period,'' as the Sun is closer to Mars, leading to higher atmospheric temperatures and increased dust levels \citep[e.g.,][]{Smith2008}. 
To take account of the possible seasonal variations, we classified the transmittance data into the following three periods and reconstructed the transit spectrum separately: the northern summer period (Ls = 30–150$^{\circ }$), the southern summer period (Ls = 210–340$^{\circ }$), and the equinox period (Ls = 340–30, 150–210$^{\circ }$).
Figure 4 displays the calculated effective heights, with the spectra shown separately for the three seasons. In this figure, the spectra are shown for the northern summer period (Ls = 30–150$^{\circ }$) in panel (a), the southern summer period (Ls = 210–340$^{\circ }$) in panel (b), and the equinox period (Ls = 340–30, 150–210$^{\circ }$) in panel (c). The color variations in each panel represent the latitudes of the measurements. This demonstrates that the spectral shape exhibits significant variability depending on latitude, even within the same seasonal period, largely reflecting latitudinal differences in dust and water-ice cloud loading.

%-----------------------------------------
%-----------------------------------------

%-----------------------------------------
\begin{figure}
\centering
\includegraphics[width=1.0\hsize]{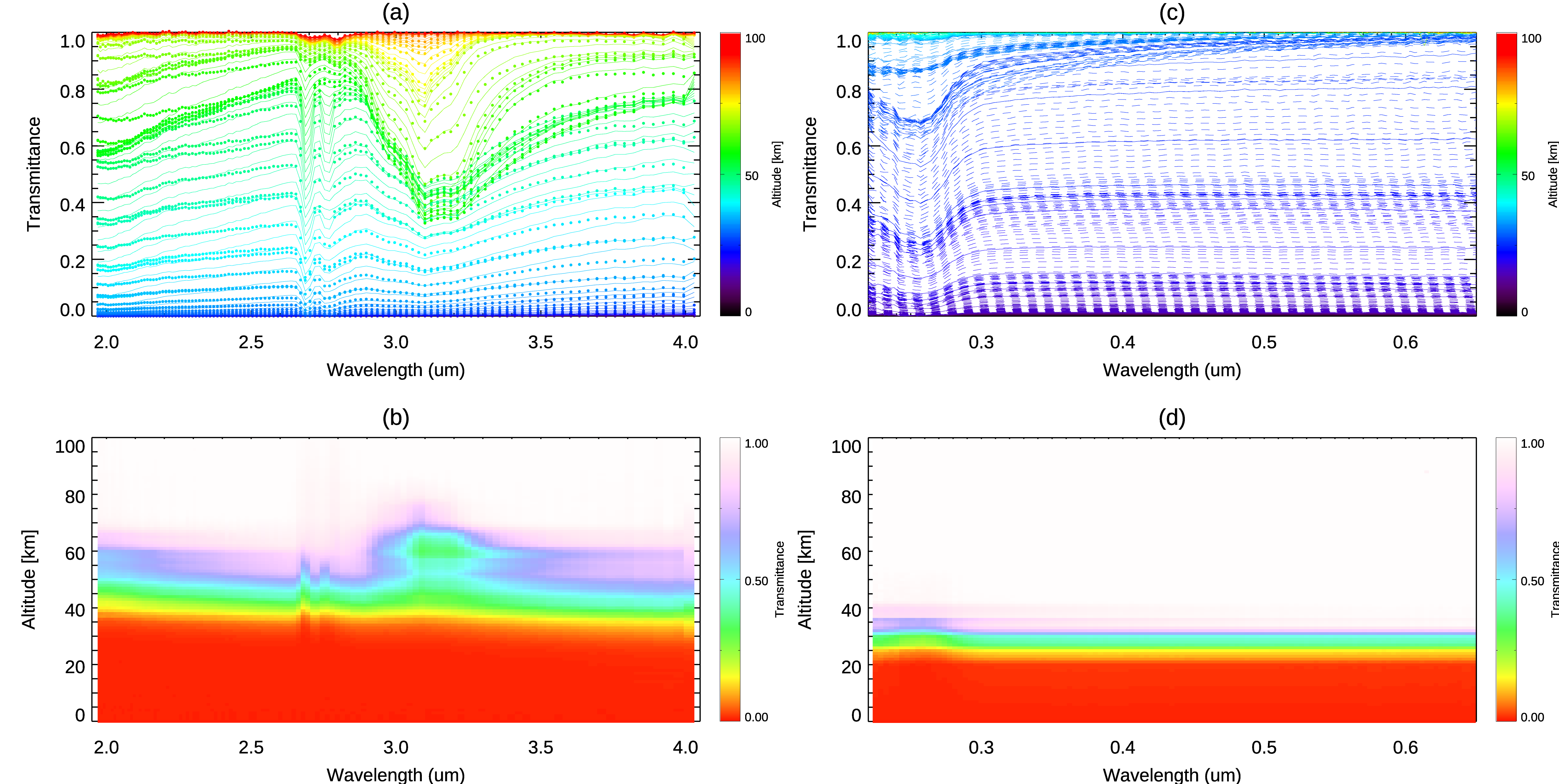}
\caption{(a,b) An example of the transmittance spectra taken with the full-scan mode of the SO channel of TGO/NOMAD. (c,d) An example of the transmittance spectra taken with the full-scan mode of the UVIS channel.}
\label{Figure 2}
\end{figure}
%-----------------------------------------

%-----------------------------------------
\begin{figure}
\centering
\includegraphics[width=0.6\hsize]{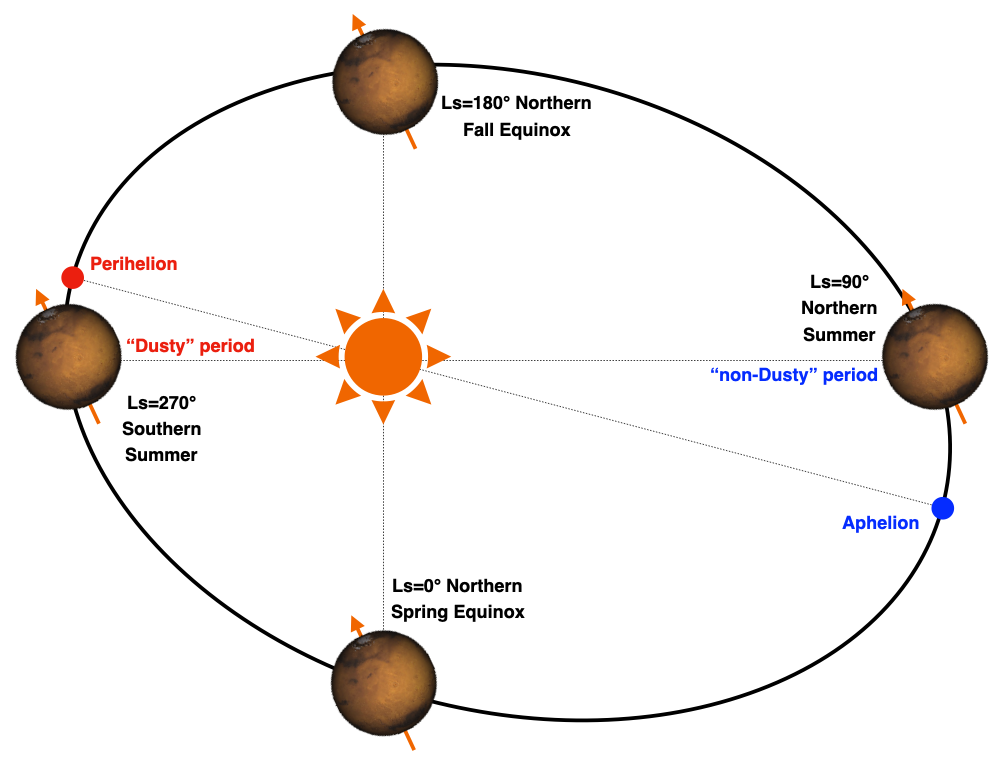}
\caption{Illustration of the relationship between the Sun and Mars throughout a Martian year. The solar longitude (Ls) is the angle between the Sun and Mars, measured from the position of the Northern hemisphere’s spring equinox, defined as Ls = 0$^{\circ }$. Thus, Ls = 90$^{\circ }$ corresponds to the northern summer solstice, Ls = 180$^{\circ }$ marks the northern autumn equinox, and Ls = 270$^{\circ }$ corresponds to the northern winter solstice.}
\label{Figure 3}
\end{figure}
%-----------------------------------------

%-----------------------------------------
\begin{figure}
\centering
\includegraphics[width=1.0\hsize]{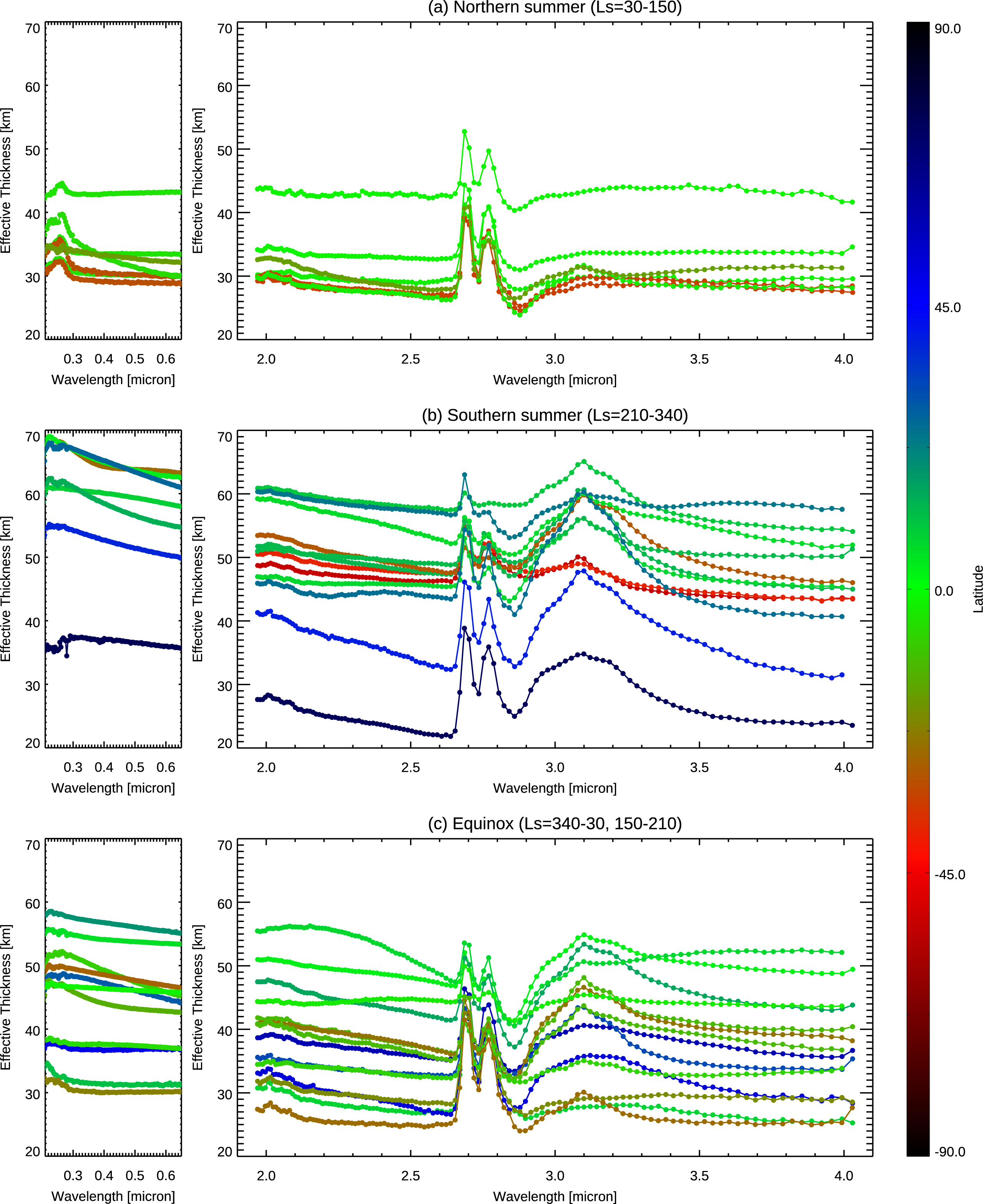}
\caption{Empirical transit spectra generated from the TGO/NOMAD Solar occultation measurements taken at (a) northern summer (Ls=30--150$^{\circ }$), (b) southern summer (Ls=210--340$^{\circ }$), and (c) equinox (Ls=340--30$^{\circ }$, 150--210$^{\circ }$). The color variation represents the latitudes of the measurements. The results obtained from the UVIS and infrared SO channels are shown in the left and right panels, respectively. }
\label{Figure 5}
\end{figure}
%-----------------------------------------

The final step of the procedure, Step 5, involves averaging the effective height spectra over latitude to construct representative empirical transit spectra. Each effective height spectrum at a given latitude, $h_{\lambda}$, is first grouped into latitude bins spanning from –90$^{\circ }$ to +90$^{\circ }$ in 30$^{\circ }$ increments. The spectra within each bin are then averaged, and finally, a global mean spectrum is obtained by averaging across all latitude bins.

\subsection{Spectral Features in the Generated Empirical Transit Spectra}

Figure 5 presents representative empirical transit spectra, averaged separately over the three different periods shown in Figure 4. As illustrated, the effective thickness of Mars' transit spectra ranges between 25 and 65 km, indicating that the atmosphere below $\sim$25 km ($\sim$0.5 mbar) is largely opaque. 
The effective thickness varies with the Martian season, being greater during dusty periods and smaller during non-dusty periods.
We note that using the NOMAD/SO spectra without averaging over the diffraction order yields a consistent baseline effective thickness (not shown), indicating that the atmosphere remains opaque below 25~km due to aerosol opacity even when the spectral resolution is increased. 

%-----------------------------------------
\begin{figure}
\centering
\includegraphics[width=1.0\hsize]{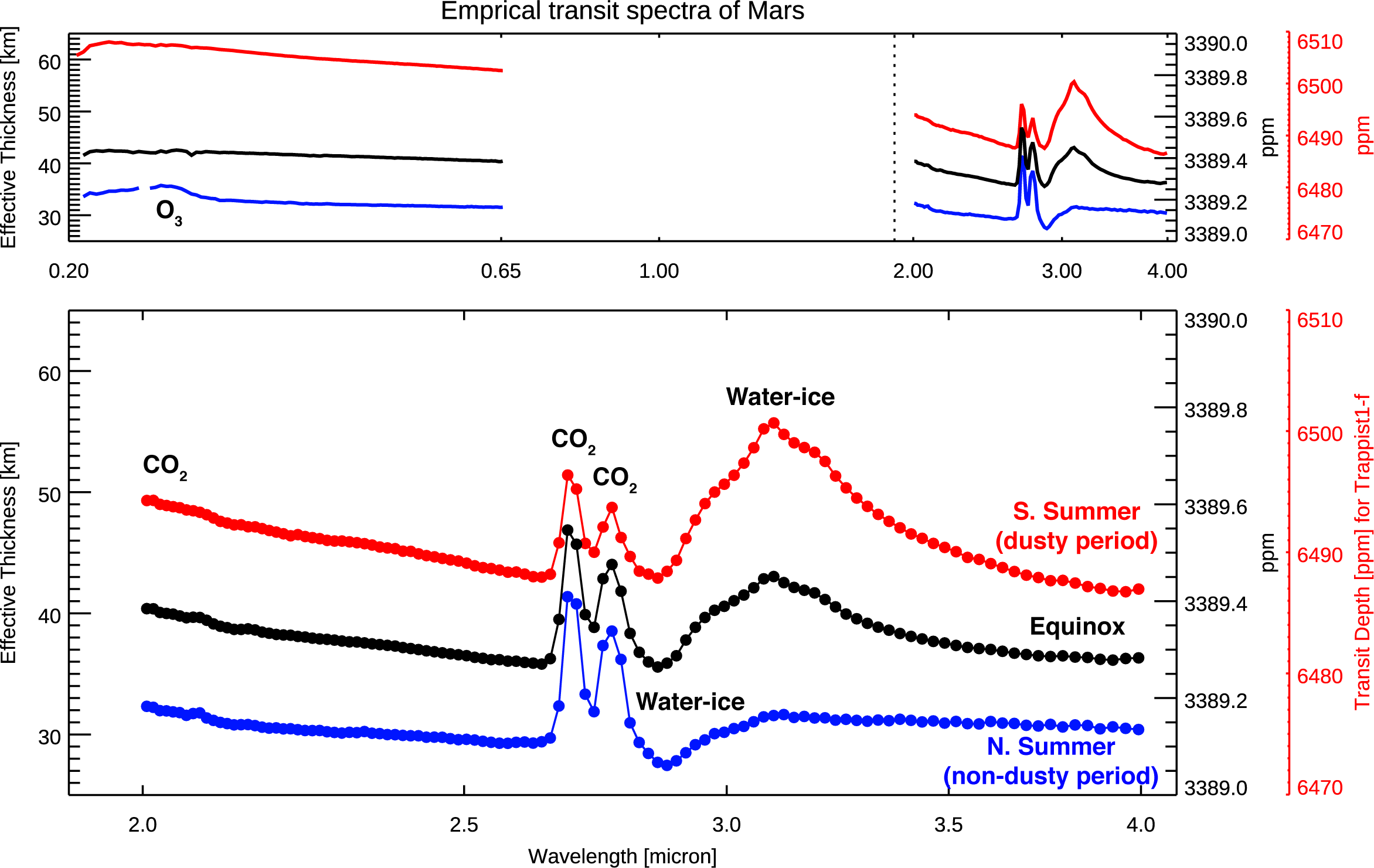}
\caption{Empirical transit spectra of Mars derived from TGO/NOMAD measurements are shown for the northern summer period (blue curve), equinox period (black curve), and southern summer period (red curve). The right y-axis indicates the transit depth for Mars (black axis), scaled to represent the case of TRAPPIST-1f.}
\label{Figure 5}
\end{figure}
%-----------------------------------------

Figure 5 highlights two distinct spectral features: CO$_{2}$ absorption at 2.7–2.8 \micron{} and \water{} ice cloud signatures %around 3.1 \micron{}
near 3 \micron{}, along with the continuum slope. 
The amplitude of the CO$_{2}$ absorption features corresponds to an effective thickness of approximately 10 km, varying with Martian seasons. This absorption is slightly weaker at dusty period. 
Weak features at 2.0–2.1 \micron{} are also due to CO$_{2}$. 

The prominent \water{} ice cloud feature is intriguing given the low abundance of water in Mars' atmosphere and the limited attention such clouds have received in previous studies.  
The feature is more prominent in the warmest ``Southern summer'' season, which may be counterintuitive given that the mesospheric clouds would form in a cold environment.  
We will further study this seasonality in Section 3.

In the UV, a weak O$_3$ feature appears near 0.25~\micron{}, and is more pronounced during Northern summer (the non-dusty season). This variation reflects the well-known seasonal cycle of Martian ozone. O$_3$ is produced from CO$_2$ photolysis and is efficiently destroyed by hydrogen radicals originating from water vapor \citep[e.g.,][]{Lefèvre2021}. During the non-dusty season, water vapor is largely confined to the lower atmosphere (below $\sim$15~km) \citep[e.g.,][]{Aoki2022}, which reduces HO$_x$-driven ozone loss and leads to higher ozone abundances \citep[e.g.,][]{Patel2021}.

In both wavelength domains, the transit spectra exhibit a spectral slope, with the effective thickness increasing toward shorter (UV) wavelengths. The spectral slope can be quantified using the power-law index of opacity, $\alpha$, defined by $\kappa \propto \lambda^{\alpha}$, which is related to the effective height through $H \alpha = \frac{dh_{\rm eff}}{d\log \lambda}$ \citep[e.g.,][]{LecavelierDesEtangs+2008}. In the 2--2.6~\micron{} range, the observed slope corresponds to $\alpha \sim -2$. This is shallower than the Rayleigh expectation, indicating that aerosols dominate the continuum opacity in this spectral region. 
On the UV, the spectral index is smaller than this. 
If the infrared and UV slopes share a common origin, the change in the spectral gradient between 0.65 and 2~\micron{} suggests characteristic aerosol particle sizes of order $\sim$1~\micron{}. 
We note, however, that the two slopes may not necessarily be explained by a single aerosol population.
Analyses combining UV and IR solar occultation data suggest that a bimodal aerosol size distribution may be required \citep{Fedorova+2014}. 
Investigating such UV-specific complexity is beyond the scope of the present work.

%______________________________________________
\section{Interpretation with informed forward modeling}
\label{s:interpretation}

In this section, the derived empirical transit spectra are quantitatively interpreted through  comparisons with modeled spectra (i.e., forward modeling). 
We focus on the infrared portion of the spectra, motivated by its relevance to potential exoplanet applications. This choice is driven not only by the success of infrared spectroscopy with JWST in characterizing atmospheric compositions, but also by the fact that promising targets for transit spectroscopy among potentially habitable planets are typically orbiting low-mass stars, which are intrinsically faint in the ultraviolet and visible wavelengths and thus subject to high photon noise in those bands.
We generated synthetic spectra using the Planetary Spectrum Generator (PSG) \citep{Villanueva2018}, an radiative transfer suite widely used for analyzing spectroscopic observations of Mars, including those from TGO/NOMAD \citep[e.g.,][]{Villanueva2021} and those from ground-based observations \citep[e.g.,][]{Aoki2024}, as well as for other Solar System bodies and exoplanets. 

\subsection{Inputs for Forward Modeling with PSG}
To compute synthetic transit spectra of Mars, we simulate observations of Mars transiting the Sun from a distance of 10 parsecs to fully emulate the geometry of a Mars transit event although the exact distance does not affect the noise-free spectra. 
PSG performs radiative transfer calculations using the line-by-line method for high spectral resolution and the correlated-k method for low spectral resolution, and we employed the latter for interpreting these spectra. The instrumental line shape was modeled as a Gaussian with a spectral resolving power ($\mathcal{R}$) of 140. This value reflects the averaging of the original NOMAD spectra—initially at $\mathcal{R}\sim $17,000—across more than 120 spectral points.

For our PSG simulations, we employ the default Martian atmospheric settings, which provide a vertical resolution of approximately 5 km in the 20–80 km altitude range.
The calculations include molecular absorptions from CO$_2$, H$_2$O, and CO, as well as extinction due to dust and \water{} ice clouds, and Rayleigh scattering. Molecular absorption coefficients were taken from the HITRAN2020 database \citep{Gordon2022}, using CO$_2$ pressure-broadening coefficients. Dust extinction coefficients were based on \citet{Wolff2009}, using CRISM data from MRO, while \water{} ice cloud extinction coefficients were derived from \citet{Warren2008}. 

The temperature structure at each season, as well as the molecular profiles, were fixed to the values predicted by the Martian Climate Database \citep[MCD,][]{Millour2018} at those instances as implemented in PSG. 
We further assume the basic properties of the dust and \water{} ice clouds. 
Namely, the representative vertical extent and particle size of the aerosols were defined a priori based on the knowledge about Mars as described below. However, their mixing ratios and the particle size of \water{} ice clouds were tuned to fit the data within the range of observed values. 
Specifically, the volume mixing ratios were varied between 0.1 and 50 ppm for both dust and \water{} ice clouds, while the effective particle size of \water{} ice clouds was adjusted within the range of 0.1 to 10 \micron{} \citep{Liuzzi+2024}.

For dust, the effective particle size is fixed at 1.5 \micron{}, and their vertical distribution is assumed to be uniform. We note that, in reality, occasional non-well-mixed vertical profiles, such as detached layers, have been suggested \citep[e.g.,][]{Heavens2011}. 

For \water{} ice clouds, we consider two types: low-altitude clouds in the upper troposphere (10–40 km) and high-altitude clouds in the mesosphere (40–70 km). The volume mixing ratio and effective particle size of the \water{} ice clouds in each layer are adjusted accordingly.
This is based on the known properties of Martian clouds.
In the Martian upper troposphere, one of the most prominent occurrences of \water{} ice clouds is the so-called “equatorial cloud belt,” which forms in specific low-latitude regions. This cloud belt develops as air cools in the ascending branch of the Hadley circulation during the northern spring–summer season, when atmospheric temperatures are relatively low due to Mars' greater distance from the Sun \citep[e.g.,][]{Montmessin2004}. The particle sizes of these stratospheric \water{} ice clouds are typically on the order of a few microns.
On the other hand, mesospheric \water{} ice clouds had been identified relatively recently \citep[e.g.,][]{Clancy+2019},
and have been systematically studied using the datasets same as what we use here, based on solar occultation measurements by TGO \citep[e.g.,][]{Liuzzi2020, Liuzzi+2024, Stolzenbach+2023, Stolzenbach+2023b}. These studies revealed that sub-micron-sized \water{} ice clouds are widely present in the mesosphere during the southern spring–summer season, when water vapor directly reaches mesospheric altitudes \citep[e.g.,][]{Aoki2022}. 
To take account of such known size difference between the stratospheric cloud and mesospheric clouds, we treat these two clouds separately, with the division at 40~km altitude.

%-----------------------------------------
\begin{figure}
\centering
\includegraphics[width=0.9\hsize]{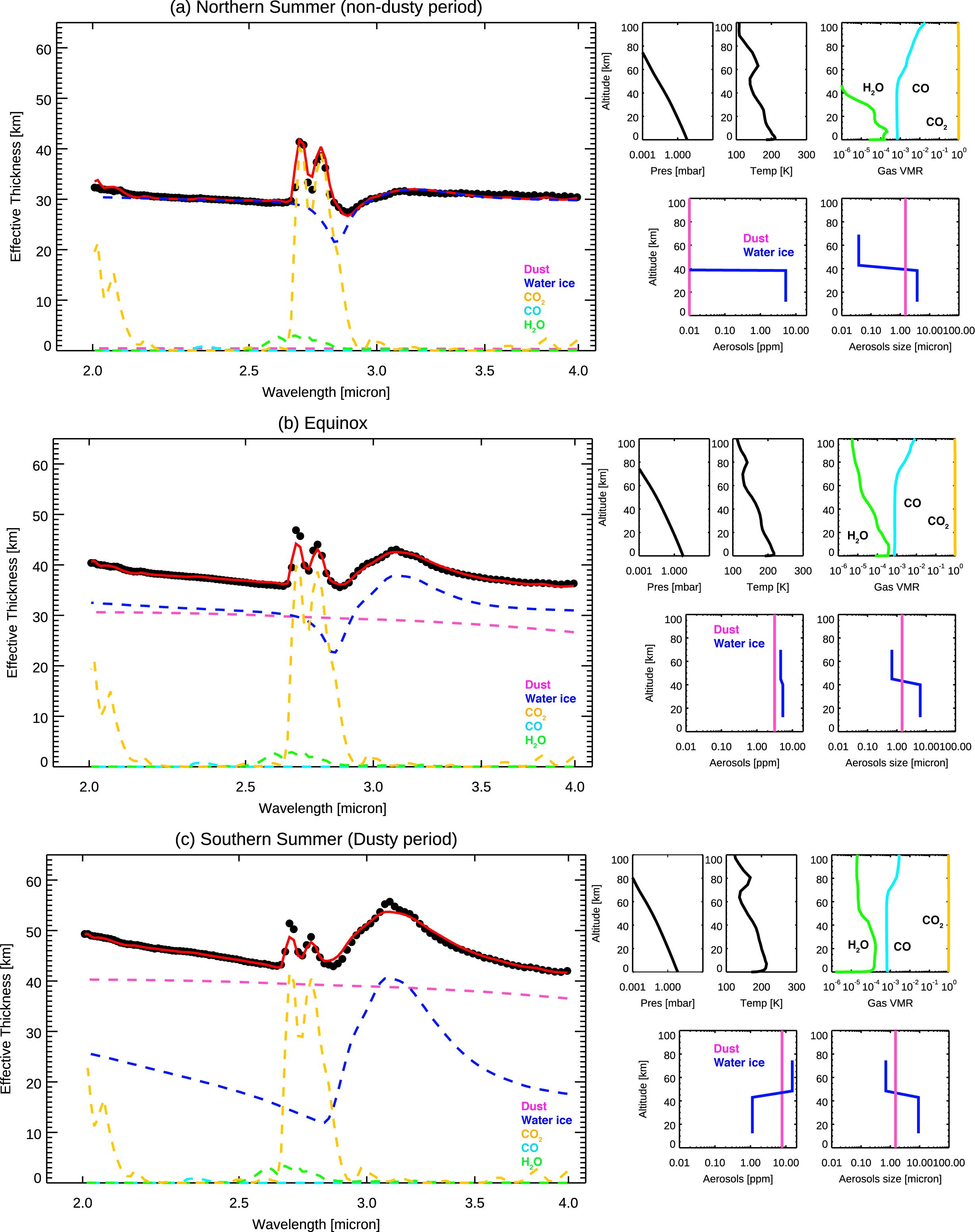}
\caption{A comparison between the empirical transit spectra of Mars, derived from TGO/NOMAD solar occultation measurements, and synthetic spectra calculated using PSG—based on a single representative atmospheric structure—is presented in the left panels for the northern summer period (a), equinox period (b), and southern summer period (c). The black curves represent the empirical transit spectra, the red curves correspond to the PSG synthetic spectra, and the other colors indicate the individual spectral contributions. The right panels show the assumed vertical profiles of pressure, temperature, gas mixing ratios (\water{}, CO, CO$_{2}$), aerosol properties (dust and \water{} ice clouds), and aerosol particle sizes.}
\label{Figure 6}
\end{figure}
%-----------------------------------------
\subsection{Comparison Between Empirical Transit Spectra and Forward Model Results}
Figure 6 compares the empirical and synthetic transit spectra, with the assumed vertical profiles shown in the right panel. This figure demonstrates that simple forward modeling, based on a single representative atmospheric structure, can effectively reproduce the observed empirical transit spectra.

The seasonal variation in CO$_{2}$ absorption at 2.7–2.8 \micron{} is successfully reproduced by the PSG simulations. The depth of the feature appears slightly weaker during periods of higher atmospheric dust loading. Since CO$_{2}$ is the dominant component of the Martian atmosphere, maintaining an almost constant volume mixing ratio of 99.6 percent throughout the year, these seasonal changes in CO$_{2}$ absorption are not due to changes in gas abundance. Instead, they are attributed to the influence of aerosols—primarily dust particles —whose mixing ratio varies in season. 
These aerosols add to the total opactiy of the atmosphere, thereby increasing the peak effective thickness at the CO$_2$ absorption band. 
In addition, they diminish the spectral signatures of CO and \water{}. Although the absorption features of \water{} and CO are centered at 2.6 \micron{} and 2.35 \micron{}, respectively (see green and light blue curves in Figure 6), these features are largely masked in the empirical transit spectra due to aerosol opacity. 
Theoretically, the spectral features of CO and \water{} become more discernible at higher spectral resolving power—for example, corresponding to an effective thickness variation of ~1 km at a resolution of R = 5000—even in the presence of dust and water ice clouds (not shown).
This underscores the critical role of dust in shaping the vertical sensitivity of transit spectra and highlights the strong seasonal variability of spectral features as a defining characteristic of Mars’ dusty atmosphere.

\begin{figure}
\centering
\includegraphics[width=0.6\hsize]{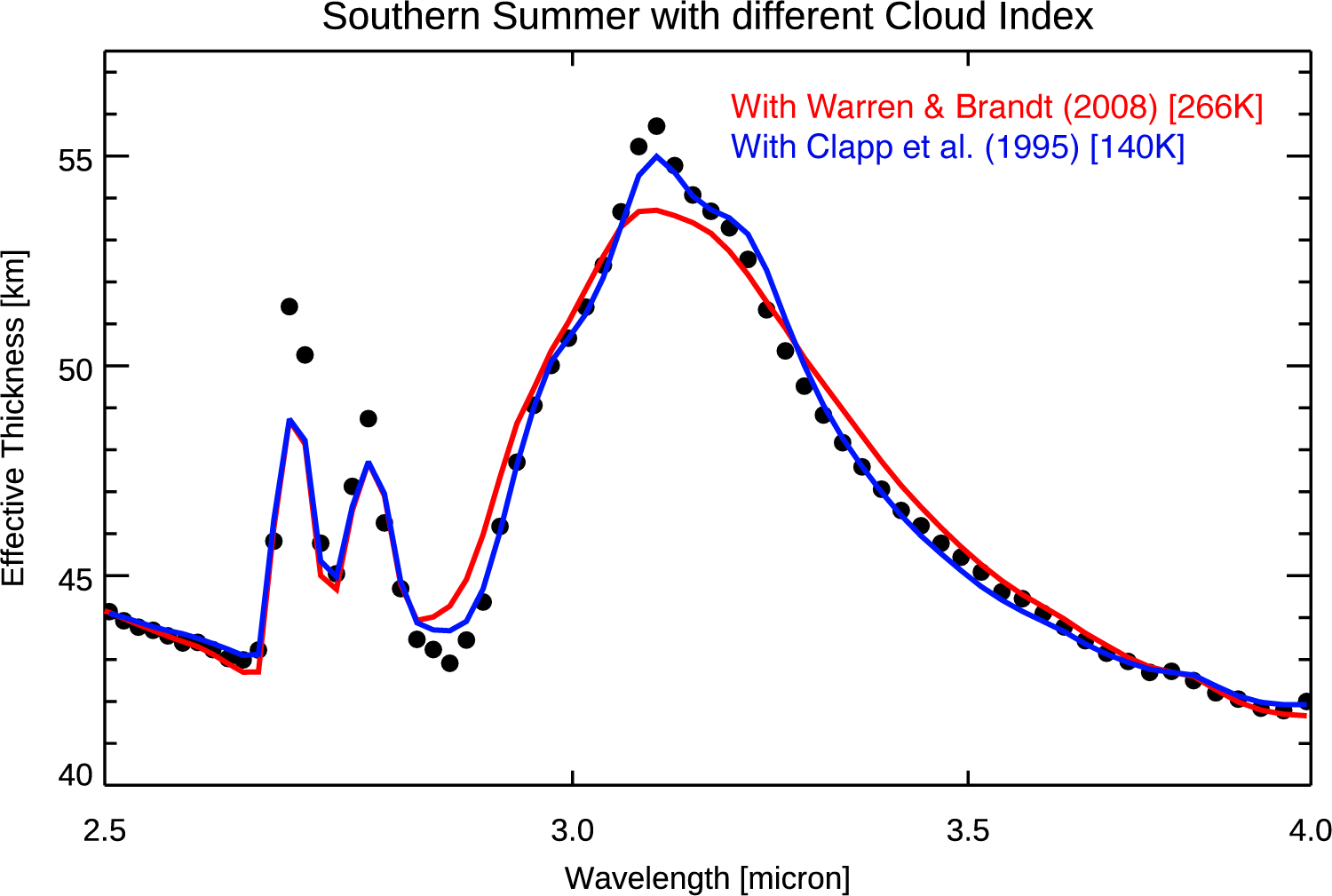}
\caption{Same as the left panel in Fig. 6c but comparison between the synthetic spectra calculated with refractive index of \water{} ice clouds by \citet{Clapp+1995} (blue) and \citet{Warren2008} (red). }
\label{fig:H2Oice_Tdep}
\end{figure}

The PSG simulation also successfully captures the seasonal variability of \water{} ice cloud signatures near 3 \micron{}. 
The reduced effective thickness around 2.9 \micron{} in the northern summer transit spectra (Fig. 6a) is attributed to the presence of larger \water{} ice particles, with radii exceeding 1 \micron{}. In contrast, the enhanced effective thickness around 3.1 \micron{} in the equinox and southern summer transit spectra (Figs. 6b, c) is attributed to the small \water{} ice particles with sizes smaller than 1~\micron{}, consistent with the sub-micron-sized mesospheric clouds on Mars.
Indeed, the presence of sharp 3.1~\micron{} feature requires the presence of sub-micron-sized particles, as we will discuss later (e.g., Figure \ref{fig:sp_reff} below).
We note that the sharp peak of the \water{} ice cloud feature at 3.1 \micron{} is not perfectly reproduced by the forward simulation (see Figure  6c). This discrepancy likely arises from the temperature dependence of the optical constants of \water{} ice.
The assumed refractive index of \water{} ice from \citet{Warren2008} is based on laboratory measurements at 266 K, which is higher than the actual temperature of the Martian mesosphere (approximately 140 K). 
A forward simulation using the refractive index measured at 140 K by \citet{Clapp+1995} provides a better match at the 3.1 \micron{} peak (Figure \ref{fig:H2Oice_Tdep}). 
However, that refractive index dataset is not used in other parts of this study due to its limited spectral coverage, only longer than 2.5 \micron{}.

%-----------------------------------------
%-----------------------------------------

%______________________________________________
\section{Potential of characterizing Mars-like exoplanets}
\label{s:exoplanet}

In this section, we aim to quantify the detectability of the characteristic features seen in Mars' transit spectra assuming it were observed as an exoplanet. 
To this end, we consider a scenario in which one of the known transiting exoplanets orbiting nearby low-mass stars, TRAPPIST-1f, has an atmospheric $T$-$P$ profile similar to that of Mars. 
TRAPPIST-1f is one of the ``golden'' targets for atmospheric characterization among temperate Earth-sized exoplanets. It receives an insolation comparable to what Mars receives from the Sun, making it an ideal testbed for our study. 
We generate mock data by adding artificial noise to the empirical transit spectra introduced in Section \ref{s:data}, and analyze it using a retrieval framework commonly employed in the study of exoplanet spectroscopy. 
As in previous section, we focus exclusively on the infrared portion of the spectra, as it is most relevant to foreseeable exoplanet observations (see Section 3).
The procedure is detailed in Section \ref{ss:exoplanet_method} and the result is presented in Section \ref{ss:exoplanet_result}. 

The reason for not assuming a Mars-twin around solar-type stars is not only that such planets have not been discovered in the solar neighborhood but also that the atmospheric signals of a Mars-twin orbiting a Sun-like star are as low as 0.1~ppm of the stellar light \citep[e.g.,][]{Fujii+2018}, which is too small to be detectable under statistic and systematic noises. 
Nevertheless, the estimate for the required precision provided below can be trivially scaled to such cases when needed.

\subsection{Method: Preparation of input data and analysis framework}
\label{ss:exoplanet_method}

\subsubsection{Input data}

We scale the empirical transit spectra of Mars to TRAPPIST-1f as follows. 
First, the effective thickness is scaled according to the scale height. Specifically, 
\begin{equation}
h_{\rm T1f} = h_{\rm Mars} \frac{H_{\rm T1f}}{H_{\rm Mars}} = h_{\rm Mars} \frac{g_{\rm Mars}}{g_{\rm T1f}} \frac{T_{\rm T1f}}{T_{\rm Mars}} = h_{\rm Mars} \frac{M_{\rm Mars}}{M_{\rm T1f}} \left( \frac{R_{\rm T1f}}{R_{\rm Mars}}\right)^2 \left( \frac{S_{\rm T1f}}{S_{\rm Mars}}\right)^{1/4} \sim 0.4 h_{\rm Mars}, 
\label{eq:scaling}
\end{equation}
where $H_X$, $g_X$, $M_X$, $R_X$, $T_X$, and $S_X$ are the scale height, surface gravity, mass, radius, equilibrium temperature, and incident flux, respectively, of planet $X$.
To obtain the right-hand side, we used the astrophysical parameters of TRAPPIST-1 as the host star and TRAPPIST-1f as the planet given by \citet{Agol+2021}, specifically $M_{\rm T1f}=1.039 M_{\oplus }$ $R_{\rm T1f}=1.045R_{\oplus} $, and $S_{\rm T1f}=0.373S_{\oplus}$ combined with $M_{\rm Mars}=0.107M_{\oplus }$, $R_{\rm Mars}=0.53R_{\oplus }$, and $S_{\rm T1f}=0.43S_{\oplus}$.
The transit depth is then calculated as $( R_{\rm T1f} + h_{\rm T1f})^2 / R_{\rm T1}^2$ ($\sim  (R_{\rm T1f}/R_{\rm T1})^2+2R_{\rm T1f}h_{\rm T1f}/R_{\rm T1}$). 
This scaling is presented on the $y2$-axis of Figure \ref{Figure 5}.

We simply impose noise levels of 3~ppm, 5~ppm, and 10~ppm per wavelength bin on the scaled signal and use the resulting data as mock observations for retrieval analysis. 
The JWST Exposure Time Calculator (ETC) estimates that a 1-hour exposure (corresponding $\sim $1 transit) at $\lambda \sim 3\,\mu $m with the NIRSpec Prism/CLEAR configuration yields a signal-to-noise ratio of $\sim $9,700 with respect to the stellar flux, for spectral elements with a width of $\Delta \lambda = 0.013\,\mu $m. 
This is equivalent to $\sim $100~ppm noise.
Given that the noise inversely scales with the square root of the photon counts, and therefore with the square root of the wavelength-bin width, this 100-ppm noise corresponds to $\sim $81~ppm noise 
for the spectral-element width of our data, $\Delta\lambda \sim 0.0212$~\micron{} at $\sim 3$~\micron{}.
By accumulating photon counts, the noise decreases to $\sim $10~ppm in 65~hours, to 5~ppm in $\sim $260~hours, or to 3~ppm in $\sim$720~hours.
The latter two are probably too long to be practical, and achieving such precisions would likely be hindered by the presence of systematic noise. 
Thus, these cases illustrate the potential capabilities of future observatories beyond JWST.

\subsubsection{Retrieval framework}

For the retrieval analysis, we used \texttt{petitRADTRANS} \citep{Molliere+2019}, a frequently used open-source code for analyzing exoplanet atmospheric data. While Section \ref{s:interpretation} takes account of the detailed vertical atmospheric structures, such complexities are unlikely to be constrained from low-spectral-resolution, low-SNR spectra without any prior knowledge of the surface environment of the planet. 
Thus, this section simply assumes a 1-dimensional atmospheric profile with constant mixing ratios of molecules and condensates, in a similar manner to the standard analysis of exoplanet data. 

For the atmospheric opacity, we considered CO$_2$, \water{} ice, and an agnostic power-law slope.
From the standpoint of being more agnostic or blind, we also searched for other typical molecules that could potentially explain the spectral features and thus be misidentified, but did not find any. Although the 3.0~\micron{} NH$_3$ band and the 3.3~\micron{} CH$_4$ band partially overlap with the 3.1~\micron{} feature, our retrievals confirm that the differences in the peak wavelengths prevent misidentification. Including either species does not improve the fit significance. Therefore, we do not consider these additional species further.

We employ the default $k$-table for the opacity of CO$_2$ based on the ExoMol line database \citep{Yurchenko+2020}, where the line profiles are collision-broadened by H$_2$. 
We used the %temperature-dependent 
opacity table for H$_2$O ice with varying particle size provided by \texttt{petitRADTRANS} based on \citet{Smith+1994}.  
The ice particles are assumed to be spheres where the opacity is calculated using the Mie theory, and the possible effects of irregular or fractal shapes of ice particles are not taken into account. 
The particle size distribution is assumed to be a log-normal function represented by the average size (effective size; $r_{\rm eff}$) and the standard deviation ($\sigma_r$), which are set as fitting parameters. 
The power-law continuum opacity is characterized by the opacity at a reference wavelength of 0.35 \micron{}, $\kappa _s$, and the power-law index, $s$, such that $\kappa (\lambda ) = \kappa_s (\lambda /0.35\,\mu {\rm m})^s$. 
The parameters for retrieval, as well as their prior distributions, are summarized in Table \ref{tbl:fitting_parameters}.

\begin{table}[h]
\begin{center}
\caption{Fitting parameters for retrievals and their priors}
\begin{tabular}{cll} \hline \hline 
Symbol  & Description & Range of flat prior \\ \hline
% $g$ & Surface gravity & Fixed & 932 cm/s$^2$  \\
$R_{\rm p}$ & Radius [$R_{\oplus }$] & $[0.993,\,1.10]$ ([0.95-1.05] $\times $ (fiducial)) \\
$T$ & Temperature [K] &	$[100,\,300]$ \\
$\log\,\kappa_s$ & log$_{10}$( continuum opacity [cm$^2$ g$^{-1}$] at 0.35 \micron{} ) & $[-2,\,0]$	\\
$s$ & power-law index for continuum opacity & $[-3,\,0]$	\\
$\log\,x_{\rm CO_2}$ & log$_{10}$( CO$_2$	mass mixing ratio ) & $[-16,\,0]$  \\
$\log\,\mu_{\rm H_2O}$ & log$_{10}$( H$_2$O(s) mass mixing ratio ) & $[-16,\,0]$ \\
$\log\,r_{\rm eff}$ & log$_{10}$( effective radius of H$_2$O(s) ) &  $[-8,\,-2]$ \\
$\log \sigma _{\rm r}$ & standard deviation of H$_2$O(s) particle radius & $[0.04,\,3.0]$ \\ \hline
\end{tabular}
\label{tbl:fitting_parameters}
\end{center}
\end{table}

%______________________________________________
\subsection{Retrieval results}
\label{ss:exoplanet_result}

%-----------------------------------------
\begin{figure}
\centering
\includegraphics[width=\hsize]{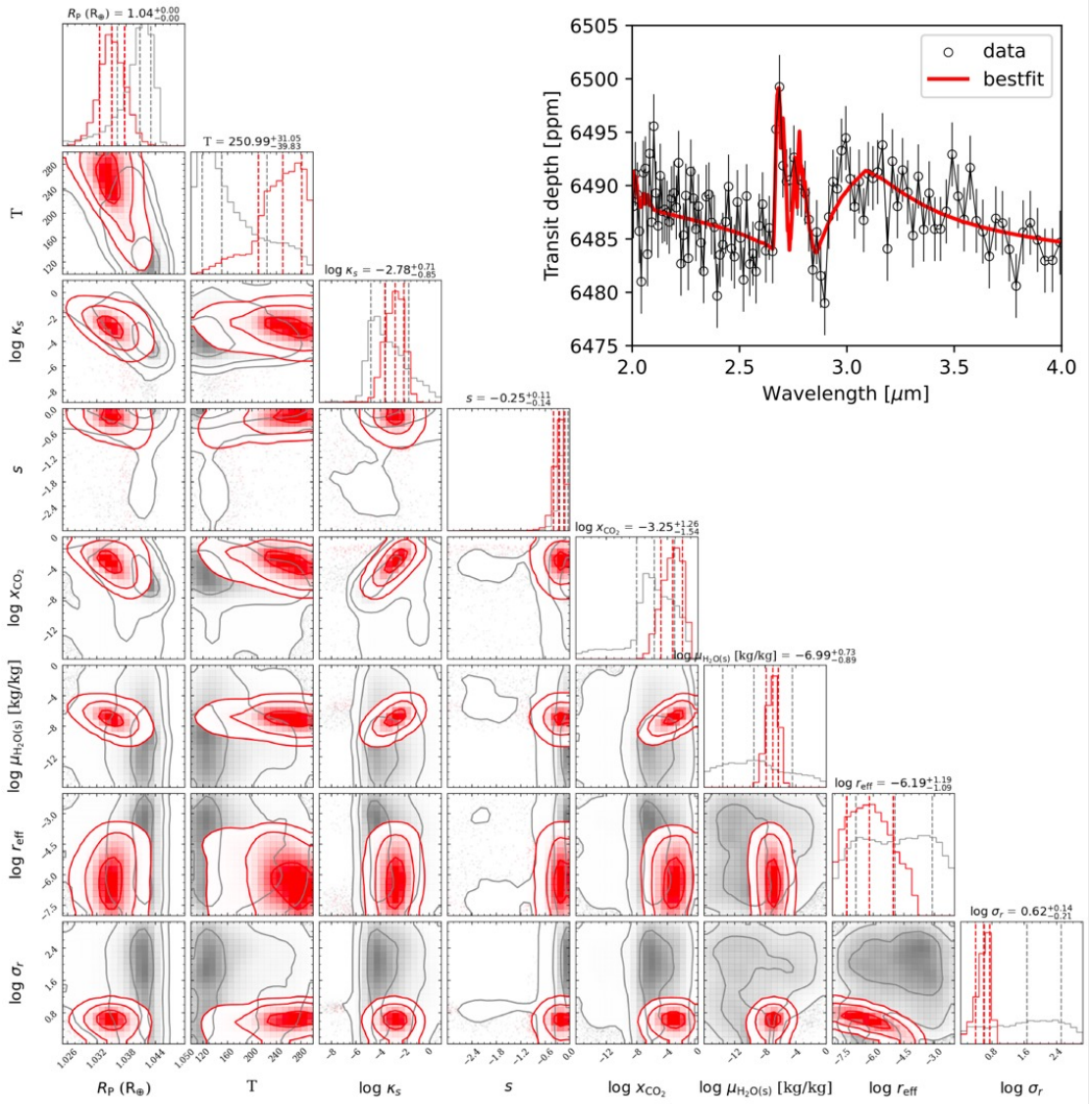}
\caption{(Left) Corner plot of the retrieval results with varying noise: 3~ppm (red) and 5~ppm (gray). (Upper right): Best-fit spectrum fitted to the data with 3~ppm noise. }
\label{fig:corner}
\end{figure}
%-----------------------------------------

The retrieval results for the empirical transit spectrum during the equinox period on Mars, assuming noise levels of 3 ppm and 5 ppm, are presented in Figure \ref{fig:corner}. 
The case with 10~ppm noise is unconstrained (not shown). 
The posterior distribution indicates that the presence of CO$_2$ and H$_2$O ice is inferred at a noise level of 3~ppm, with a flat continuum ($s\sim 0$). 
To quantify the significance of the CO$_2$ and H$_2$O ice detections, we performed additional retrievals without CO$_2$ and/or \water{} ice, and compared the Bayesian evidences, with variations in the noise realization. 
The Bayes factor comparing the CO$_2$+continuum model to the continuum-only model is $> 10^2$, while that comparing the full model (CO$_2$+H$_2$O+continuum) to the CO$_2$+continuum model is also $> 10^2$. These results indicate a “decisive” preference for the full model, according to Jeffreys’ scale \citep{Jeffreys1939}.

%-----------------------------------------
\begin{figure}
\centering
\includegraphics[width=0.5\hsize]{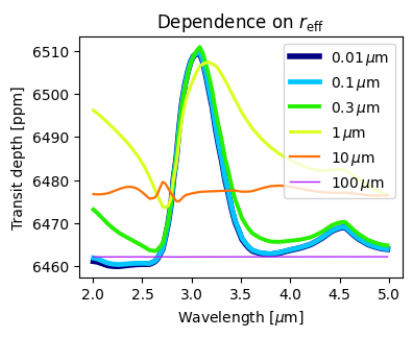}
\caption{Dependence of the 3.1 $\mu $m H$_2$O ice feature on the effective radius of H$_2$O ice clouds. The CO$_2$ feature and slope is removed. The standard deviation of the radius distribution, assumed to be log-normal, is assumed to be $\ln(\sigma_r) = 0.1$. 
%H$_2$O ice assumed are: $\mu_{\rm H_2O}=10^{-6}$ kg/kg and $\ln(\sigma_r) = 0.4$. 
Planetary parameters are set to the values of TRAPPIST-1f. }
\label{fig:sp_reff}
\end{figure}
%-----------------------------------------

The effective radius of \water{} ice, as well as the width of the size distribution, is well-constrained at a noise level of 3~ppm.  
Specifically, the particle size larger than 1~\micron{} is strongly disfavored. 
The upper limit arises from the known fact that the 3~\micron{} feature is substantially suppressed at $r_{\rm eff} \gtrsim 1$~\micron{}. 
This behavior is illustrated in Figure \ref{fig:sp_reff}, which shows the dependence of the transit spectrum on the particle size of \water{} ice cloud.
In general, the absorption feature appears only in the Rayleigh regime, where $2\pi r_{\rm eff}/\lambda < 1$.

The number density of \water{} ice cloud particles at the transit radius at $\lambda $ = 3.1~\micron{}, denoted by $n_{\rm 0}$, can be estimated from the condition that the slant optical depth is order of unity. 
Letting $\kappa_{\rm ice}$ represent the mass extinction coefficient of \water{} ice, this condition can be expressed as \citep[e.g.,][]{Fortney2005}
\begin{equation}
\kappa_{\rm ice} \frac{4\pi r^3  \rho_{\rm ice}n_0}{3}  \sqrt{ 2\pi R_{\rm p} H } \left( \sim \tau_{\rm ice,\,nadir} \sqrt{\frac{2\pi R_{\rm p}}{H}}\right) \sim 1, \label{eq:optdepth_slant}
\end{equation}
where $\rho _{\rm ice}$ and $r$ are the density and radius of \water{} ice cloud particles, while $\tau_{\rm ice,\,nadir} $ represents the vertical optical depth. 
Substituting representative values of $\kappa_{\rm ice} \sim 10^4$~cm$^2$\,g$^{-1}$ at 3.1~\micron{}, $\rho_{\rm ice} \sim 10^3$~kg m$^{-3}$, $R_{\rm p} \sim R_{\oplus}$, and $H \sim 10^4$ m yields
\begin{equation}
n_0 \sim 10^9 \, \left( \frac{r}{0.1\,{\rm \mu m}} \right)^{-3} \,[{\rm m}^{-3}]
\end{equation}

However, the mass {\it mixing ratio} of \water{} ice clouds remains unconstrained.
This is because the pressure level (or equivalently, the density of the ambient atmosphere) at which the absorption bands originate is not well determined. 
This represents the known degeneracy between the reference pressure level and the molecular mixing ratios \citep[e.g.,][]{BennekeSeager2012,Heng+2017}.
Specifically, a small mixing ratio at high pressure and a large mixing ratio at low pressure can yield a similar number density $n_0$, and thus produce comparable absorption depths. 

Similarly, the mixing ratio of CO$_2$ is unconstrained, and the solution with 100\% CO$_2$ is not particularly favored. 
Meanwhile, the {\it relative} abundances of multiple species are constrained by the relative strengths of their spectral features. 
The posterior distribution of $x_{\rm CO_2}$ and $X_{\rm H_2O}$ (Figure \ref{fig:corner}) suggests a dependence of $x_{\rm CO_2} \sim (X_{\rm H_2O,\,(s)}/10^{-5.5})^{2}$.
The power-law index of $2$ likely reflects the scaling of collisionally broadened CO$_2$ opacity, which is proportional to $n_{\rm CO_2}p \propto x_{\rm CO_2}p^2$, where $n_{i}$ is the number density of species $i$ and $p$ is the atmospheric pressure. 
In contrast, the \water{} ice opacity scales as $n_{\rm H_2O} \propto X_{\rm H_2O}p$. 
Maintaining a constant ratio of these opacities across varying reference pressure levels leads to the above relation. 
Note that the upper limit of $x_{\rm CO_2}$ is unity, and correspondingly, the upper limit of $\mu_{\rm H_2O}$ is $\sim 10^{-5.5}$; This solution retrieves values consistent with those for Mars in our forward models presented in the middle panel of Figure 6.

%%%%%%%%%%%%%%%%%
\begin{table}[h]
\begin{center}
\caption{Best-fit parameters for the retievals with fixed $x_{\rm CO_2}$}
\begin{tabular}{ccccc} \hline \hline 
& \multicolumn{4}{c}{$x_{\rm CO_2}$} \\ \cline{2-5}
Symbol  & 1 & $10^{-2}$ & $10^{-4}$ & $10^{-6}$ \\ \hline
$R_{\rm p}$ [$R_{\oplus}$] & 1.406 & 1.0375 & 	1.0378 & 1.0396 \\
$T$ [K] & 158 & 172 & 209 & 232  \\
$\log\,\kappa_s$ & -1.9 & -2.5 & -3.37 & -4.4 \\
$s$ & -0.35 & -0.11 & -0.098 & -0.095 \\
$\log\,\mu_{\rm H_2O}$ & -5.4 & -6.5 & -7.6 & -8.7 \\
$\log\,r_{\rm eff}$ [cm] & -8.0 & -5.8 & -5.6 & -5.2 \\
$\log \sigma _{\rm r}$ & 0.83 & 0.55 & 0.51 & 0.45 \\ \hline
\end{tabular}
\label{tbl:bestfit_fixedCO2}
\end{center}
\end{table}
%%%%%%%%%%%%%%%%%

\begin{figure}
\centering
\includegraphics[width=0.5\hsize]{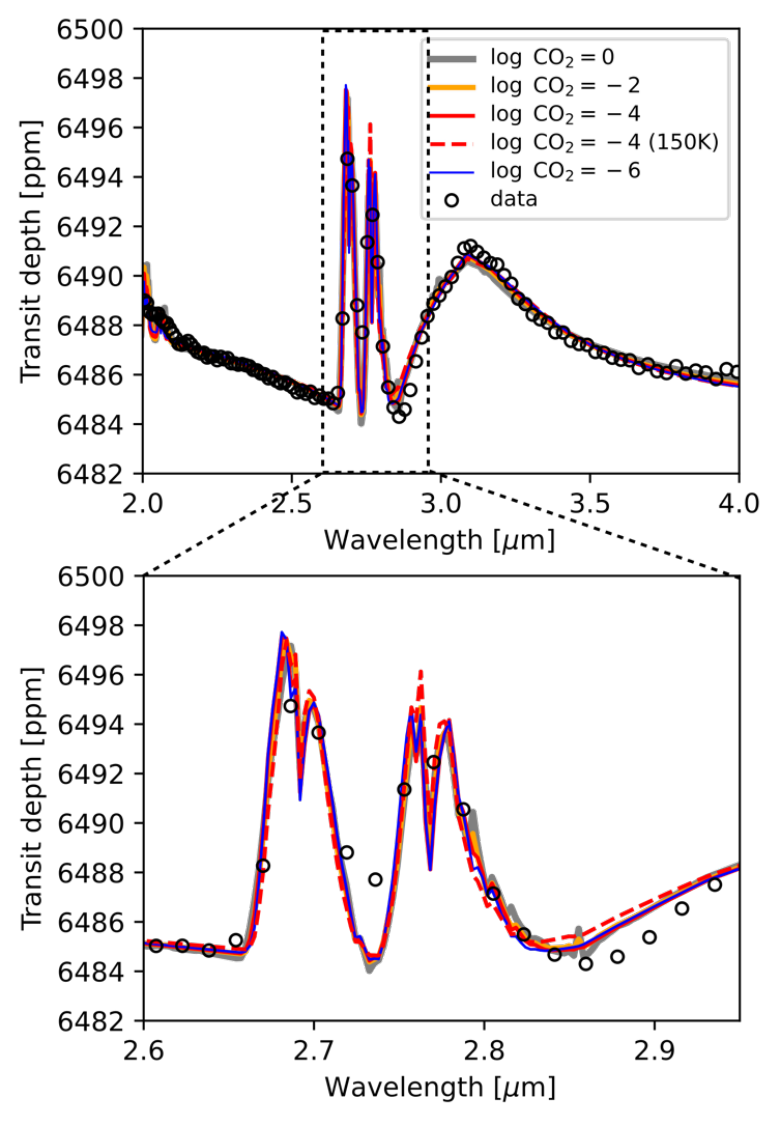}
\caption{The best-fit solutions for the noise-free transit spectra of ``Equinox'' with fixed CO$_2$ mass mixing ratios to $10^{-6}$, $10^{-2}$, and $1$. }
%$\chi^2$ of red, green, and blue lines are: 6.65, 2.89, and 3.24, respectively. }
\label{fig:spec_varying_xCO2}
\end{figure}

To investigate this degeneracy in more detail, best-fit spectra assuming fixed CO$_2$ mixing ratios (1, $10^{-2}$, $10^{-4}$, and $10^{-6}$) are obtained and shown in Figure~\ref{fig:spec_varying_xCO2}. 
The corresponding best-fit parameters are listed in Table~\ref{tbl:bestfit_fixedCO2}. 
While the detailed spectral shapes vary slightly, all models reproduce the observations equally well given the current level of data accuracy. 
Table~\ref{tbl:bestfit_fixedCO2} clearly shows that synchronized changes in the CO$_2$ and H$_2$O mixing ratios, as well as in $\kappa_s$, act to preserve the relative strengths of the spectral features. 
The increasing temperatures at lower CO$_2$ mixing ratios likely serve to reduce the discrepancy around 2.75--2.95~\micron{}, as demonstrated by the red dashed line in Figure~\ref{fig:spec_varying_xCO2}, which corresponds to the best-fit solution at $x_{\rm CO_2} = 10^{-4}$ with a fixed temperature of 150~K. 
The planetary radius is adjusted to match the absolute level of the transit depth. 
Small molecular mixing ratios tend to be compensated by a larger planetary radius, except in the case of $x_{\rm CO_2} = 1$, where the increased mean molecular weight reduces the scale height and thus requires a larger planetary radius at the reference pressure. 
Constraining the predominance of CO$_2$ in the atmosphere will require data with higher accuracy and spectral resolution. 

%-----------------------------------------
%-----------------------------------------

As mentioned above, the required noise level as low as 3~ppm is likely too demanding even for the state-of-the-art observations with JWST. 
It should be noted that the challenge is due in part to the presence of dust particles, rather than the direct effect of the low atmospheric pressure (as might be initially speculated). 
A simple dust-free CO$_2$ 100\% atmosphere with a Mars-like surface pressure (6~mbar) would have more than 3 times larger signal (see Figure 6), enabling the detection of CO$_2$ with higher noise level. 

While the result for TRAPPIST-1f is shown here, the expected signal (and hence the required noise level for detection) for other TRAPPIST-1 planets can be scaled by $R_X h_X$ with equation (\ref{eq:scaling}) for $h_X$. 
Specifically, based on the planetary parameters given in \citet{Agol+2021} ($M_{\rm T1d}=0.388M_{\oplus}$, $R_{\rm T1d}=0.788R_{\oplus}$, $S_{\rm T1d}=1.115S_{\oplus}$, $M_{\rm T1e}=0.692M_{\oplus}$, $R_{\rm T1e}=0.92R_{\oplus}$, $S_{\rm T1d}=0.646S_{\oplus}$) and assuming equal albedos, the signals of TRAPPIST-1d and TRAPPIST-1e are $\sim $1.5 times and $\sim $1.2 times that of TRAPPIST-1f.

%______________________________________________
%\section{Summary and discussions}
\section{Discussions}
\label{s:icefeature}

\subsection{Temporal variability of molecular features}

One of the distinct features in the empirical transit spectra of Mars is the temporal variation in the depth of the CO$_{2}$ absorption band at 2.7–2.8 \micron{} driven by the planet's dust cycle. 
Could similar variability in transit spectra be expected for rocky planets orbiting M-type stars? 
A difference from Mars is that exoplanets are observed only at a fixed orbital phase--that is, the same season. 
Thus, the seasonal variations demonstrated in this study may not appear directly relevant to such targets. 
However, multi-year observations of Mars have revealed that dust activity can vary significantly from year to year, even during the same season. 
For example, large-scale, planet-encircling dust storms occur irregularly, approximately once every five Martian years \citep[e.g.,][]{Montabone+2015}.
If similar intermittent dust storms occur on rocky exoplanets, they may contribute to temporal variability in transit spectra. 

It should be noted that the processes that causes the time variability of transit spectra does not necessarily involve dust storms. 
Significant temporal variations in aerosol extinction above the cloud layer have also been observed on Venus \citep{Wilquet+2012}. 
A Mars-like dusty thin atmosphere could be identified by further constraining the dust composition to silicates, for example, through the 10 \micron{} silicate absorption feature \citep[e.g.,][]{Ruff+2002,Toon+2019}. 

\subsection{3.1~$\mu $m \water{} ice feature}

The presence of a \water{} ice signature in transit spectra is intriguing, since \water{} ice on Mars is not typically recognized as one of the most prominent spectroscopic features of the planet. 
As discussed in Section \ref{s:interpretation}, this feature is attributed to sub-micron-sized mesospheric \water{} ice clouds. 
Do similar small \water{} cloud particles form on rocky exoplanets in other systems?
Put differently, what conditions on exoplanets allow such small cloud particles to form and build up sufficient optical thickness?
Understanding the conditions under which small \water{} cloud particles can form is important because it directly informs what we can infer about an exoplanet when the 3.1~\micron{} feature is detected.
Insights into these questions can be gained from the formation processes and environmental conditions of analogous clouds on Solar System planets. 
Therefore, in what follows, we review the current understanding of \water{} ice clouds on terrestrial planets in the Solar System, and discuss what the detection of such a feature would imply in the context of exoplanet characterization. 

Earth also has mesospheric clouds, called noctilucent clouds (NLCs) or polar mesospheric clouds (PMCs). 
They occur around $\sim$1~Pa, a pressure similar to that of Mars' mesosphere but corresponding to altitudes of 80–85~km, in the summer high latitudes. 
The typical particle size of Earth's mesospheric \water{} clouds is also $<0.1$~\micron{} \citep[e.g.,][]{Rusch+1991}, small enough to exhibit the 3.1~\micron{} characteristics in principle. 
However, empirical transit spectra of Earth do not show such a feature \citep{Doshi+2022}. 
The cause is uncertain, but possibly due to the small optical depth as well as localization in space and time. 
There have been no reports of \water{} ice clouds observed in the Venusian atmosphere.
Past observations suggested that the atmospheric temperatures at the cloud layer that consists primarily of sulfuric acid droplets (45-70~km) through the aerosol layer (up to 90~km) \citep{Titov18} are too warm to allow for the formation of ice clouds. 
However, recent studies by \cite{Murray23} have provided new insights;  based on the temperature profiles obtained by the Solar Occultation in the InfraRed (SOIR) instrument onboard the Venus Express spacecraft \citep{Mahieux15, Mahieux23}, they noted that temperatures at higher altitudes, around 120~km, could fall to as low as 120~K. At such low temperatures, the formation of ice clouds becomes possible. 
They suggested that nanometer-scale amorphous solid water (ASW) particles could form across all latitudes under these conditions, and that CO$_2$ ice particles might condense on ASW surfaces at high latitudes, where temperatures are even lower. 
Such mesospheric clouds on Venus have not yet been observed, and their existence remains a subject for future investigation.

Formation of mesospheric clouds on both Mars and Earth requires 
the partial pressure of water vapor reaching the saturation vapor pressure, which depends on the water vapor profiles and the temperature structure. 
The more limited extent of Earth's mesospheric clouds compared to Mars' may be attributed to the presence of a temperature minimum near the 0.1 bar % 100~hPa 
level, which constrains the water vapor abundance below the saturation pressure at that pressure level \citep[``cold trap'', e.g.,][]{Catling+2017}. 
Due to this bottle neck in the water vapor transport, along with the relatively warm stratospheric temperature due to O$_3$, the atmosphere above the temperature minimum tends to remain under-saturated. 
This is also the case during the northern summer season on Mars, when relatively low atmospheric temperatures confine water vapor to altitudes below approximately 15–20 km \citep[e.g.,][]{Aoki2022}. This altitude is commonly referred to as the ``hygropause''. 
In contrast, mesospheric \water{} clouds on Mars form during the southern summer season when water vapor is efficiently transported upward through the heated lower atmosphere \citep[e.g.,][]{Aoki2022}.
Thus, the prominent presence of mesospheric \water{} clouds may reflect key aspects of the underlying atmospheric structure, in particular, the absence of cold trap or hygropause.

Furthermore, studies on the mesospheric clouds on Mars and Earth have highlighted the importance of the availability of cloud condensation nuclei (CCN). 
The formation PMCs on Earth likely involves heterogeneous nucleation on dust grains produced by meteoric ablation \citep[e.g.,][]{Tanaka+2022}. 
Similarly, theoretical studies suggest that Martian mesospheric \water{} ice clouds preferentially form via heterogeneous nucleation, either on dust particles lofted from the lower atmosphere \citep[e.g.,][]{Maattanen+2005}, or on exogenous dust grains originating from meteoric ablation \citep[e.g.,][]{Hartwick+2019}. 
Accordingly, the detection of mesospheric clouds on exoplanets may point to analogous processes that supply CCN in their upper atmospheres, such as the influx of micrometeoroids or the vertical transport of dust.

It should be noted that \water{} ice clouds can also form below the mesosphere. Indeed, Earth has \water{} ice clouds in the upper troposphere and stratosphere. 
Clouds in the upper troposphere, which include familiar types of clouds such as cirrus, cirrostratus, and cirrocumulus, form when moist air ascends and cools. 
Stratospheric \water{} ice clouds are rarer and typically form over polar regions during winter known as Polar Stratospheric Cloud (PSC), when the air is substantially cooled. 
The 3.1~\micron{} feature from these clouds is again absent in empirical transit spectra of Earth \citep{Doshi+2022}. 
This is  likely due (at least in part) to the particle sizes. 
Typical particle sizes of clouds in the upper troposphere and stratosphere are tens of microns \citep[e.g.,][]{Liou1986} and 2–10~\micron{} \citep[e.g.,][]{Dye1992}, respectively. 
These sizes are too large to clearly exhibit the 3.1~\micron{} feature (Figure~\ref{fig:sp_reff}), and such clouds contribute to the transit spectra primarily as gray opacity. 
In the water-rich environment of Earth's lower atmosphere, cloud particles can grow in a relatively short timescale, until they become large enough to even more rapidly grow through coalescence and/or fall 
\citep[e.g.,][]{Rossow1978}.

Whether or not exoplanets develop \water{} clouds with a sufficient slant optical depth at 3.1~\micron{} depends ultimately on the interplay among water vapor transport, heat transport, and transport of CCN via atmospheric dynamics—processes that are difficult to accurately model. 
\citet{Fauchez+2019} who used a general circulation model (GCM) to simulate the atmospheres of TRAPPIST-1 planets suggested the presence of \water{} ice clouds at the terminator with particle sizes of approximately 2~\micron{}. 
However, cloud particle sizes depend on both the number density of CCN and the details of atmospheric circulation, introducing significant modeling uncertainties. 
Similarly, the formation of mesospheric \water{} ice clouds on exoplanets is influenced by the availability of CCN, the water vapor content, and the local thermal structure—often modulated by atmospheric wave activity—making their occurrence probability uncertain. 
To the best of our knowledge, the existence of mesospheric clouds on exoplanets has not been explicitly addressed in the literature. Whether, and under what conditions, such \water{} ice clouds form and retain small particle sizes remains an open question. 

Nevertheless, it should be noted if such sub-micron-sized \water{} ice clouds do form, the 3.1~\micron{} \water{} ice absorption feature is substantially more stronger than the low-resolution \water{} {\it vapor} bands for a given water mass mixing ratio. 
Therefore, it would serve as a valuable compositional clue in future atmospheric characterization of habitable-zone exoplanets.

\section{Summary}
This study analyzes the empirical transit spectra of Mars, reconstructed from solar occultation measurements by the NOMAD instrument onboard the ExoMars Trace Gas Orbiter, to investigate how a Mars-like surface environment on exoplanets would be seen in transit observations. 
The results show that Mars' lower atmosphere is largely opaque below $\sim$25 km ($\sim$0.5 mbar) due to the presence of dust and \water{} ice clouds, while the 2.7--2.8 \micron{} CO$_2$ absorption band and distinct 3.1~\micron{} H$_2$O ice features are identified. 
These spectral features exhibit marked seasonal variability: during dusty periods, the 2.7–2.8~\micron{} CO$_2$ absorption weakens, and the H$_2$O ice feature at 3.1~\micron{} becomes more prominent. 
Forward modeling with PSG shows that the empirical spectra are well reproduced when vertical atmospheric structures based on previous observational studies are incorporated. Aerosols, particularly dust and fine ice particles, strongly influence both the continuum slope and the amplitude of spectral features. 

The 3.1~\micron{} H$_2$O feature is intriguing given the low abundance of water vapor on Mars. 
This is attributed to the sub-micron-sized mesospheric H$_2$O clouds formed above 40~km in altitude ($\sim$0.1 mbar). 
This spectral feature of \water{} is sensitive to small mass mixing ratio of water, and could be a sensitive indicator of the presence of water in cold, arid atmospheres like that of Mars, where water {\it vapor} absorption cannot be detected. 
The large spatial extent of mesospheric water clouds with small ice particles as observed for Mars (and has not been observed on Earth) is likely related to certain properties of Mars, including no prominent cold trap and airborne dust particles that efficiently heat the lower atmosphere. 
These findings highlight the diagnostic potential of aerosol- and cloud-related features for characterizing the atmospheric environments of terrestrial exoplanets. 
However, it should also be noted that uniquely identifying a planet's climate state from transmission spectra over a limited wavelength range remains a significant challenge. Such an inference likely requires multi-faceted characterization using various observational techniques, with transmission spectral properties serving as one piece of evidence.

To evaluate the detectability of such features in an exoplanetary context, we simulate transit observations of a Mars-like atmosphere scaled to TRAPPIST-1f and perform retrieval analysis. Our results indicate that both CO$_2$ and H$_2$O ice features can be detected at precisions better than $\sim$3 ppm. 
However, absolute abundance estimates are limited by uncertainties in the reconstructed transit spectra, coupled with degeneracies involving planetary radius and reference pressure. 
The required precision exceeds current technological capabilities, demonstrating the challenge of characterizing dusty atmospheres like Mars.

Our analysis adds to the limited sample of empirical transit spectra of well-studied solar system planets---Earth \citep{MacdonaldCowan2019,Doshi+2022} and Titan \citep{Robinson2014}. Given the difficulty of accurately modeling three-dimensional atmospheric properties, particularly those involving aerosols and dust, these empirical spectra serve as valuable benchmarks for future exoplanet investigations.

%______________________________________________
\begin{acknowledgments}
We greatly thank Kazumasa Ohno for the insightful discussions on the cloud properties and their impact on transit spectra. We also thank Tatsuya Yoshida for valuable discussions on the implications of Martian dust for the planet’s evolutionary history.

SA was supported by JSPS KAKENHI Grant Number 24K21565, 22K03709, 22H05151, 22H00164, and 22KK0044.
YF was supported by JSPS KAKENHI Grant Number 18K13601. 
GLV was supported by NASA’s Mars Program Office under “GSFC participation in the TGO/NOMAD Investigation of Trace Gases on Mars” and by NASA's SEEC Exoplanetary collaboration.

The NOMAD experiment is led by the Royal Belgian Institute for Space Aeronomy (IASB-BIRA) with co-PI teams from Spain (IAA-CSIC), Italy (INAF-IAPS) and the United Kingdom (Open University). This project acknowledges funding by: the Belgian Science Policy Office (BELSPO) with the financial and contractual coordination by the ESA Prodex Office (PEA 4000103401, 4000121493, 4000140753, 4000140863); by the Spanish Ministry of Science and Innovation (MCIU) and European funds (grants PGC2018-101836-B-I00 and ESP2017-87143-R; MINECO/FEDER), from the Severo Ochoa (CEX2021-001131-S) and from MCIN/AEI/10.13039/501100011033 (grants PID2022-137579NB-I00, RTI2018-100920-J-I00 and PID2022-141216NB-I00); by the UK Space Agency (grants ST/V002295/1, ST/V005332/1, ST/X006549/1, ST/Y000234/1 and ST/R003025/1); and by the Italian Space Agency (grant 2018-2-HH.0). 
\end{acknowledgments}

%% To help institutions obtain information on the effectiveness of their 
%% telescopes the AAS Journals has created a group of keywords for telescope 
%% facilities.
%
%% Following the acknowledgments section, use the following syntax and the
%% \facility{} or \facilities{} macros to list the keywords of facilities used 
%% in the research for the paper.  Each keyword is check against the master 
%% list during copy editing.  Individual instruments can be provided in 
%% parentheses, after the keyword, but they are not verified.

\vspace{5mm}

%% Similar to \facility{}, there is the optional \software command to allow 
%% authors a place to specify which programs were used during the creation of 
%% the manuscript. Authors should list each code and include either a
%% citation or url to the code inside ()s when available.

\software{\texttt{petitRADTRANS} \citep{Molliere+2019}, \texttt{Planetary Spectrum Generator} \citep{Villanueva2018}}

%% Appendix material should be preceded with a single \appendix command.
%% There should be a \section command for each appendix. Mark appendix
%% subsections with the same markup you use in the main body of the paper.

%% Each Appendix (indicated with \section) will be lettered A, B, C, etc.
%% The equation counter will reset when it encounters the \appendix
%% command and will number appendix equations (A1), (A2), etc. The
%% Figure and Table counter will not reset.

%\appendix

%\section{Appendix information}

%Appendices can be broken into separate sections just like in the main text.

\bibliography{1_bib_MarsTrans}{}
\bibliographystyle{aasjournal}

%% This command is needed to show the entire author+affiliation list when
%% the collaboration and author truncation commands are used.  It has to
%% go at the end of the manuscript.
%\allauthors

%% Include this line if you are using the \added, \replaced, \deleted
%% commands to see a summary list of all changes at the end of the article.
%\listofchanges

\end{document}